\documentclass[cits]{PoS}
\usepackage[utf8]{inputenc}
\usepackage[T1]{fontenc}
\usepackage{slashed}
\usepackage{times}

\usepackage{amssymb,amsfonts,amsmath,amsthm}
\usepackage{dsfont,bbm}
\usepackage{dcolumn}

\usepackage{epsf}
\usepackage{graphicx}
\usepackage[caption=false]{subfig}
\usepackage{dsfont}
\usepackage{slashed}
\usepackage[active]{srcltx}
\newcommand{\ms}{\mskip 1.5mu}

\newcommand{\gpdH}{H}
\newcommand{\gpdE}{E}

\newcommand{\gpdF}{F}
\newcommand{\gpdtH}{\widetilde{H}}
\newcommand{\gpdtE}{\widetilde{E}}

\newcommand{\cffH}{{\mathcal H}}
\newcommand{\cffE}{{\mathcal E}}

\newcommand{\cffF}{{\mathcal F}}
\newcommand{\cfftH}{\widetilde{\mathcal H}}
\newcommand{\cfftE}{\widetilde{\mathcal E}}

\newcommand{\cffHbmp}{{\mathfrak H}}
\newcommand{\cffEbmp}{{\mathfrak E}}

\newcommand{\cffFbmp}{{\mathfrak F}}
\newcommand{\cfftHbmp}{\widetilde{\mathfrak H}}
\newcommand{\cfftEbmp}{\widetilde{\mathfrak E}}

\newcommand{\C}{T}

\newcommand{\xB}{x_{\rm B}}
\newcommand{\Q}{Q}

\newcommand{\GeV}{{\rm GeV}}

\newcommand{\GK}{{\it GK12 }}

\def\II{\hbox{{1}\kern-.25em\hbox{l}}}

\title{Resolving kinematic ambiguities in QCD predictions for Deeply Virtual Compton Scattering}
\ShortTitle{Resolving kinematic ambiguities in QCD predictions for Deeply Virtual Compton Scattering}
\author{\speaker{V. M. Braun}$\ms{}^a$, A. N. Manashov$\ms{}^{a,b}$, D. M{\"u}ller$\ms{}^c$ and B. Pirnay$\ms{}^a$ 
        \\
$^a$ Institut f\"ur Theoretische Physik, Universit\"at Regensburg, D-93040 Regensburg, Germany \\
$^b$ Department of Theoretical Physics,  St.-Petersburg University, 199034 St.-Petersburg, Russia\\
$^c$ Institut f\"ur Theoretische Physik II, Ruhr-Universit\"at Bochum,D-44780 Bochum, Germany  \\ \vspace*{-0.2cm} \\
E-mail: \email{vladimir.braun@physik.uni-regensburg.de}}

\abstract{The existing QCD predictions for the Deeply Virtual Compton Scattering (DVCS) depend on the convention
          used for the skewedness parameter and on the reference frame used to define helicity amplitudes.
          These ambiquities are formally power-suppressed but numerically significant. They are cancelled by 
          finite-$t$ and target mass corrections that have been calculated recently to the $1/Q^2$ accuracy. 
          It turns out that these corrections can be minimized, at least for unpolarized observables, 
          by choosing a specific reference frame where longitudinal directions are defined by the photon     
          momenta.}

\FullConference{XXII. International Workshop on Deep-Inelastic Scattering and Related Subjects \\
                 28 April - 2 May 2014 \\
                 Warsaw, Poland}

\begin{document}

%\section{Introduction}

Deeply Virtual Compton Scattering (DVCS) is the simplest process that gives access to
generalized parton distributions (GPDs) and is receiving a lot of
attention. The existing experimental results come from DESY (HERA H1, ZEUS and HERMES) and Jefferson Lab (Hall A and CLAS) 
and many more measurements are planned after the Jefferson Lab $12$~GeV upgrade and at COMPASS-II at CERN. 
The standard theoretical framework is based on collinear factorization which is proven in QCD to the
leading power accuracy in the photon virtuality $Q$.
This approach, commonly referred to as the leading twist (LT) approximation, 
appears to be sufficient to describe main features of the available data raising the
hope that a fully quantitative description is within reach. The future data will have much higher 
statistics and allow one to extract at least some GPDs with controllable precision.

The LT approximation is, however, incomplete and in fact convention-dependent.
It is well known that the LT DVCS amplitudes do not satisfy electromagnetic Ward identities.
The Lorentz (translation) invariance is violated as well: The results depend on the frame of reference chosen
to define the skewedness parameter and the helicity amplitudes.
The required symmetries are restored by a subset of higher-twist, i.e power-suppressed, corrections
$\sim t/Q^2$ and $\sim m^2/Q^2$ 
that  can be called kinematic as they are expressed in terms of the same GPDs that enter the
LT amplitudes, i.e. they do not involve new nonperturbative input. 
One expects that the subset of kinematical power corrections is factorizable
for arbitrary twist.
%
%The required symmetries are restored by contributions that are formally suppressed by  powers
%of $1/Q$, dubbed higher-twist corrections.
%Such power corrections can be called kinematic as they are expressed in terms of the same GPDs that enter the
%LT amplitudes, i.e. they do not involve new nonperturbative input. Their role, from the theory point of view,
%is to restore exact symmetries that are broken in the LT approximation and make the
%calculation unambiguous. By this reason one can expect that the subset of kinematical power corrections is factorizable
%for arbitrary twist.
%
The structure of kinematic 
%$t/Q^2$ and target mass $m^2/Q^2$ 
corrections turns out to be  nontrivial and was understood only 
recently~\cite{Braun:2011zr,Braun:2011dg,Braun:2012bg,Braun:2012hq}. 
The first detailed study of their impact on various DVCS observables is presented in \cite{Braun:2014sta}. 

%\section{Reference frame dependence}

The reason why the LT approximation is intrinsically ambiguous
is that in the DVCS kinematics the four-momenta of the initial and final photons and protons do not lie in
one plane. Hence the distinction of longitudinal and transverse directions is convention-dependent. In the Bjorken
high-energy limit this is a $1/Q$ effect.
The freedom to redefine large `plus' parton momenta by adding smaller transverse components has two consequences.
First, the relation of the skewedness parameter $\xi$ appearing as an argument in GPDs
to the Bjorken variable $x_B$ may involve power suppressed contributions.
Second, such a redefinition generally leads to excitation of the subleading photon helicity-flip amplitudes.
Any attempt to compare the calculations with and without kinematic power corrections must start with specifying the
precise conventions, i.e.~the \emph{definition} of what is meant by `leading-twist' to the power accuracy.
This is an important point that is often overlooked in phenomenological studies.

The common wisdom is that at leading order (LO) and the LT level
there are four Compton form factors (CFFs) $\cffF \in \{\cffH,\cffE,\cfftH,\cfftE\}$ that are given by
convolution integrals of GPDs $\gpdF\in \{\gpdH,\gpdE,\gpdtH,\gpdtE\}$ over the momentum fraction $x$ with simple coefficient functions,
\begin{align}
\cffF & \stackrel{\rm LO}{=} \sum_q e_q^2 \int_{-1}^1\!\! \frac{dx}{2\xi}\,
\C_{0}\biggl(\!\frac{\xi+x- i\epsilon}{2(\xi-i\epsilon\!)}\biggr)
 \gpdF^{q^{(+)}}(x,\xi,t)\,
  ~\stackrel{\rm LO}{\equiv}  \C_0\!\circledast\!\gpdF\,, && \C_0(u) = \frac1{1-u}
\label{DVCS-LO}
\end{align}
with an obvious correspondence
$\cffH \leftrightarrow \gpdH\,,\;\;\cffE \leftrightarrow \gpdE\,,\;\; \cfftH \leftrightarrow \gpdtH\,,\;\;\cfftE \leftrightarrow \gpdtE\,.$
Here and below $\gpdF \equiv \gpdF^{q^{(+)}}(x,\xi,t)$ are the $C = +1$ combinations of the GPDs defined with the 
established conventions, 
%e.g., given in \cite{Diehl:2003ny},  
and we have introduced a notation `$\circledast$' for the (normalized) convolution integral, including the sum over the quark flavors.

If the QCD calculation is done to the ${1/Q^2}$ accuracy, the following complications occur and must be taken into account:
a) The skewedness parameter $\xi$ must be \emph{defined} with a power accuracy
\begin{align}
   \xi \to \xi(x_B,t,Q^2) = \frac{x_B}{2-x_B} + \mathcal{O}(1/Q^2)\,,
\end{align}
b) The CFFs must be \emph{defined} through a certain decomposition of the DVCS tensor. 
        The LO CFFs (\ref{DVCS-LO}) are
        recovered as the scaling limit of the helicity-conserving CFFs, that is
\begin{align}
 {}\hspace*{0.6cm}\mathcal{F}_{++} &= \C_0\!\circledast\!\gpdF\Big|_{\xi\to \xi(x_B,t,Q^2)}  + \mathcal{O}(1/Q, 1/Q^2),
\end{align}
 where the expression for the $\mathcal{O}(1/Q, 1/Q^2)$ addenda depends both on the chosen form factor decomposition
and on the convention used for the skewedness parameter.
c) There are eight more CFFs $\mathcal{F}_{0+},\mathcal{F}_{-+} $ corresponding to photon helicity flip transitions that must be taken into account
        in the same approximation.
The existing freedom in definitions is related to the
choice of the reference frame in which one performs the calculation.
It is important to realize
that the corresponding ambiguities only cancel at the level of physical observables.

One possibility is to use a certain generalization of the standard DIS reference
frame where the initial photon and proton momenta form the longitudinal plane.
In the Belitsky, M{\"u}ller and Ji (BMJ) reference frame~\cite{Belitsky:2001ns,Belitsky:2010jw, Belitsky:2012ch} the nucleon target is at rest, 
$p_1^\mu=(m,0,0,0)$, and the incoming photon momentum is specified as
$
q_1^\mu=({Q}/{\gamma}) \bigl(1,0,0,-\sqrt{1+\gamma^2}\bigr)$ with $\gamma \equiv \epsilon^{\rm BMJ} = {2 m \xB}/{Q}$.
The polarization vectors of the initial photon are defined as
$\epsilon_1^\mu(0)=({1}/{\gamma})\bigl(-\sqrt{1+\gamma^2},0,0,1\bigr)$ and
$\epsilon_1^\mu(\pm)=({e^{\mp i\phi}}/{\sqrt{2}})\bigl(0,1,\pm i,0\bigr)$,
where the phase is given by the azimuthal angle $\phi$ of the final state nucleon.
BMJ employ the KM convention for the skewedness variable $\xi(\xB, t,Q^2) =  \xi_{\rm KM} = {\xB}/(2-\xB)$,
used by Kumeri{\v c}ki and M{\"u}ller in global DVCS fits~\cite{Kumericki:2009uq,Kumericki:2013br}.
A complete parametrization of the Compton tensor in terms of CFFs in this frame was proposed in
Ref.~\cite{Belitsky:2001ns}. Starting from this parametrization, the electroproduction
cross section has been calculated~\cite{Belitsky:2010jw, Belitsky:2012ch} (BMJ) for all possible polarization options
of the initial electron and nucleon.

In contrast to this traditional approach,
Braun, Manashov and Pirnay (BMP)~\cite{Braun:2012bg,Braun:2012hq} define the longitudinal plane as spanned by the \emph{two photon} momenta
$q_1$ and $q_2$. 
For this choice the momentum transfer to the target $\Delta = q_1-q_2$ is
purely longitudinal and both --- initial and final state --- protons have the same nonvanishing transverse momentum $P_{\perp}$,
such that
$
  |\xi P_\perp|^2 = (1/4)(1-\xi^2) (t_{\rm min}-t)\,,\quad  t_{\rm min} = - 4 m^2 \xi^2/(1-\xi^2)\,,
$
where $\xi = \xi_{\rm BMP}$ is the BMP skewedness parameter defined
with respect to the real (final state) photon momentum $q_2^2=0$:
$
  \xi_{\rm BMP}  =  \xB(1+t/Q^2)/[2-\xB(1-t/Q^2)].
$
The main advantage of this choice is that polarization of both the initial and the final photon can be described using the 
same polarization vectors, see Appendix~A in Ref.~\cite{Braun:2014sta}, that allows to obtain a comparatively simple
Lorentz-decomposition of the Compton tensor in terms of the CFFs.   

The BMP CFFs $\cffFbmp \in\{\cffHbmp,\cffEbmp, \cfftHbmp, \cfftEbmp\}$ are, however, different from their
BMJ analogues
$\cffF\in \{\cffH,\cffE,\cfftH,\cfftE\}$. The relation can easily be worked out~\cite{Braun:2014sta}:
\begin{align}
\label{CFF-F2F}
\cffF_{\pm+} =  \cffFbmp_{\pm+} + \frac{\varkappa}{2}\Big[\cffFbmp_{++}+\cffFbmp_{-+}\Big]- \varkappa_0\, \cffFbmp_{0+},
&&
\cffF_{0+} =  - \left(1+\varkappa\right) \cffFbmp_{0+} + \varkappa_0\Big[\cffFbmp_{++} + \cffFbmp_{-+}\Big]
\end{align}
with an obvious correspondence $\cffH \leftrightarrow \cffHbmp$, etc.
Explicit expressions for the kinematic factors 
$\varkappa_0 = \mathcal{O}(1/Q) $ and $\varkappa = \mathcal{O}(1/Q^2)$  
are given in Eq.~(48) in~\cite{Braun:2014sta}.
Using the exact transformation formulas from the
BMP to the BMJ basis, Eq. (\ref{CFF-F2F}), one can calculate physical observables 
from the expressions given in Ref.~\cite{Belitsky:2012ch}.
In this way the results are the same as the corresponding results 
which one would obtain by a direct calculation by means of the original
BMP parametrization.

%Starting from 
Within the BMJ conventions, the LT approximation to LO accuracy
can be summarized as:
\begin{align}
  \text{LT} \equiv \text{LT}_{\rm KM} &: ~\begin{cases}
         \cffF_{++}=\C_0\!\circledast\!F, & \cffF_{0+}= 0
          \\
         \cffF_{-+}=0,  & \xi = \xi_{\rm KM}
        \end{cases}
\label{KM-convention}
\end{align}
i.e.~the BMJ helicity-conserving CFF is calculated in the LO approximation using
$\xi_{\rm KM} = x_B/(2-x_B)$ for the skewedness parameter and the other CFFs are put to zero.
This ansatz~\cite{Kumericki:2013br} in practical terms is not very different from the VGG convention used by Guidal, 
and also the convention used by Kroll, Moutarde and Sabatie in \cite{Kroll:2012sm}.

Starting instead from the BMP framework, the analogous LT and LO approximation, rewritten in terms of the 
BMJ CFFs using the transformation rules in (\ref{CFF-F2F}), reads
\begin{align}
  \text{LT}_{\rm BMP} &: ~\begin{cases}
         \cffF_{++}=\left(\!1+\frac{\varkappa}{2}\!\right) \cffFbmp_{++}\,,
         %\C_0\!\circledast\!F,
         & \cffF_{0+}= \varkappa_0\, \cffFbmp_{++} %\varkappa_0 \C_0\!\circledast\!F,
          \\[1mm]
         \cffF_{-+}=\frac{\varkappa}{2}\cffFbmp_{++}, % \C_0\!\circledast\!F,
         & \xi = \xi_{\rm BMP}\,.
        \end{cases}
\label{BMP-convention2}
\end{align}
The two LT ans\"atze in Eq.~(\ref{KM-convention}) and Eq.~(\ref{BMP-convention2}) are both perfectly
legitimate. Their difference reveals that both the distinction between helicity-conserving and
helicity-flip CFFs, and the expression for skewedness parameter in terms of kinematic invariants,
depend to power $1/Q$ accuracy on the reference frame.
The resulting ambiguity is quite large because, first, the kinematic factors $\varkappa_0$ and $\varkappa$ are sizable
despite of being power-suppressed. For example, for $-t/\Q^2  \simeq 1/4$ one obtains $\varkappa/2 \sim 1/3$.
Second, $\xi_{\rm BMP} < \xi_{\rm KM}$, for practical purposes one can approximate $\xi_{\rm BMP} \approx (1+t/Q^2)\xi_{\rm KM}$ for $\xB \le 0.4$.
Thus generally $F(\xi_{\rm BMP},\xi_{\rm BMP}) >  F(\xi_{\rm KM},\xi_{\rm KM})$ if the GPDs have Regge behavior, although this effect is moderated
for larger $t$ by the slope of the Regge-trajectory.

This ambiguity is resolved by adding kinematic power corrections to the Compton amplitude that correspond to contributions of 
higher-twist operators of special type, obtained from the LT operators by adding total derivatives~\cite{Braun:2011zr,Braun:2011dg}.
The same contributions restore electromagnetic gauge invariance and translation invariance of the results. To the $\mathcal{O}(1/Q^2)$
accuracy one obtains~\cite{Braun:2012hq,Braun:2014sta}, e.g. for $\cffHbmp_{ab}$ CFFs in the BMP basis:
\begin{align}
\cffHbmp_{++} &=
  {T}_0\circledast H  + \frac{t}{Q^2}\Big[-\frac{1}{2}{T}_0 + {T}_1
+ 2 \xi  \mathbb{D}_\xi\,{T}_2 \Big]\circledast H
+ \frac{2t}{Q^2} \xi^2 \partial_\xi \xi  {T}_2 \circledast M \,,
\notag\\
 \cffHbmp_{0+}&= - \frac{4 |\xi P_\perp|}{\sqrt{2}Q}\Big[
  \xi \partial_\xi {T}_1 \circledast H   + \frac{t}{Q^2} \partial_\xi \xi\,{T}_1 \circledast M \Big]
- \frac{t}{\sqrt{2}Q |P_\perp|} {T}_1 \circledast \Big[\xi\, M -\widetilde{H} \Big]\,,
 \\
 \cffHbmp_{-+}&=\frac{4|\xi P_\perp|^2 }{Q^2}\Big[
\xi \partial^2_\xi \xi \,T_1^{(+)} \circledast H  + \frac{t}{Q^2} \partial^2_\xi \xi^2\, T_1^{(+)} \circledast M \Big]
%\notag\\ &
+
\frac{2t}{Q^2} \xi
\Big[\xi\partial_\xi\xi \,T_1^{(+)} \circledast M + \partial_\xi\,\xi\,T_1 \circledast \widetilde H\Big],
\notag
\end{align}
where $M= H+E$, $\mathbb{D}_\xi = \partial_\xi + (2/t) {|\xi P_\perp|^2}\partial^2 _\xi \xi$ and $\xi\equiv\xi_{\rm BMP}$.
The new coefficient functions appearing in these expressions are defined as
\begin{align}
\C^{(+)}_{1}(u) =\,\frac{(1-2u)\ln(1-u)}{u},
&&
\C_{1}(u) = -\frac{\ln(1-u)}{u},
&&
\C_2(u) =&\, \frac{{\rm Li}_2(1)-{\rm Li}_2(u)}{1-u} + \frac{\ln(1-u)}{2u}.
\label{T2-coef}
\end{align}
The LT$_{\rm BMP}$ approximation takes into account the first term ${T}_0\circledast H$ in $\cffHbmp_{++}$
and neglects the addenda (and the helicity-flip CFFs).
The expressions for $\cffEbmp_{ab}$, $\cfftHbmp_{ab}$, $\cfftEbmp_{ab}$ CFFs are similar~\cite{Braun:2012hq,Braun:2014sta}.

\nopagebreak

In Ref.~\cite{Braun:2014sta} we have carried out a detailed comparison of the  LT$_{\rm KM}$ and  LT$_{\rm BMP}$ approximations vs.
the complete calculation to the $1/Q^2$ accuracy  for several key
DVCS observables for unpolarized and longitudinally polarized targets. 
%
%%%%%%%%%%%%%%%%%%%%%%%%%%%%%%%%%%%%%%%%%%%%%%%%%%%%%%%%%%%
\begin{figure}[t]
\begin{center}
\includegraphics[width=15cm]{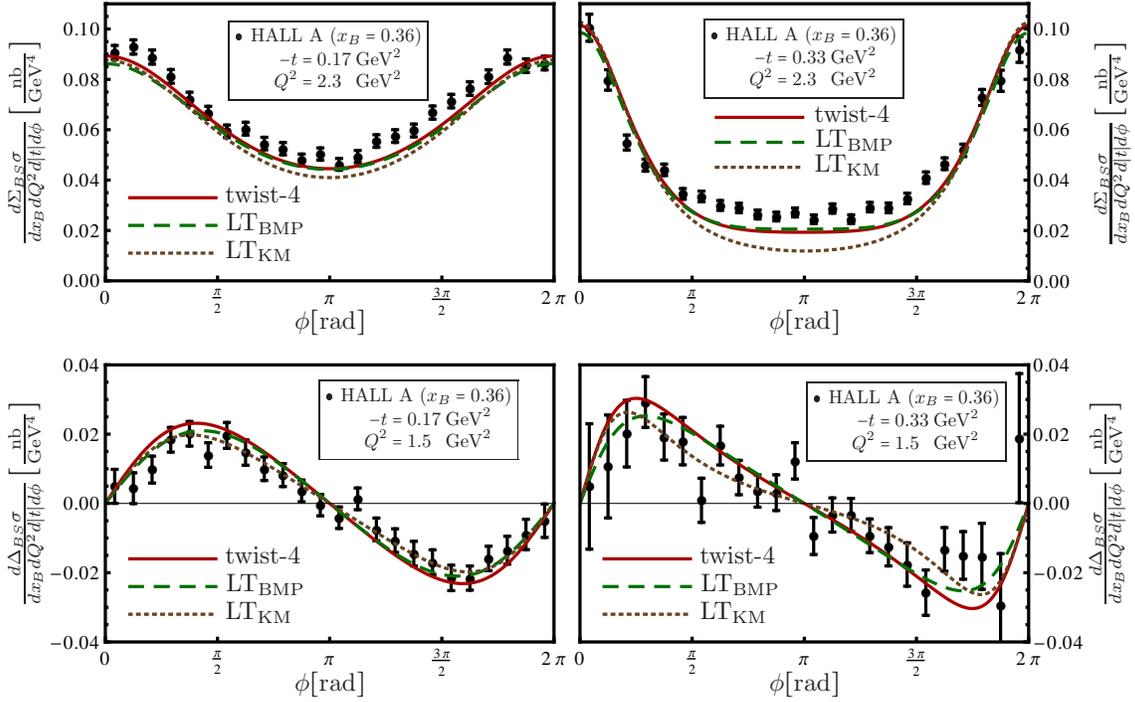} %\includegraphics[width=8cm]{plots/fig5b}
\caption{The unpolarized cross section [upper panels]
 and electron helicity dependent cross section difference [lower panels] from 
 HALL A collaboration \cite{Munoz_Camacho:2006hx} vs.~\GK GPD model predictions.
% in the
% LT=LT$_{\rm KM}$ and LT$_{\rm BMP}$ approximation [dotted and dashed curves, respectively]
% and with full account of kinematic $1/Q^2$ power corrections [solid curves].
 }
\label{Fig:HALLA}
\end{center}
\end{figure}
%%%%%%%%%%%%%%%%%%%%%%%%%%%%%%%%%%%%%%%%%%%%%%%%%%%%%%%%%%%%
%
As an example, we show in Fig.~\ref{Fig:HALLA} [upper panels] the HALL A collaboration data~\cite{Munoz_Camacho:2006hx} 
for the unpolarized cross section. Their description using popular GPD models is widely regarded as  challenging.
The data are compared with the QCD calculation using the \GK GPD model in three different approximations:
LT$_{\rm KM}$ (dotted curves), LT$_{\rm BMP}$ (dashed curves),  and with the full account of kinematic twist-four effects (solid curves).
The BH squared term is calculated using the formula set from~\cite{Belitsky:2001ns}.
Because of this contribution, the differences of the predictions of the unpolarized cross section in
different models or approximations are washed out.
%In the conventional LT$_{\rm KM}$ framework, the \GK GPD model underestimates the  data slightly
%for the smallest $-t$ value and strongly for the large $-t$. 
Changing $\text{LT}_{\rm KM} \to \text{LT}_{\rm BMP}$ produces relative large enhancement of both the DVCS cross section and the 
interference term and the prediction becomes closer to the data, whereas the remaining
kinematical twist corrections are hardly visible.
Thus, for this observable, the $\text{LT}_{\rm BMP}$ approximation alone captures the main part of the total kinematic power correction.

The electron helicity dependent cross section difference is shown in  Fig.~\ref{Fig:HALLA} in the two lower panels.
For $t=-0.33\,\GeV^2$ the differences in the three predictions are clearly visible and affect significantly the shape of the 
$\phi$-distribution. Having in mind the experimental errors, all of the predictions are, nevertheless, compatible with the data.
\begin{figure}[t]
{}\includegraphics[width=0.5\textwidth]{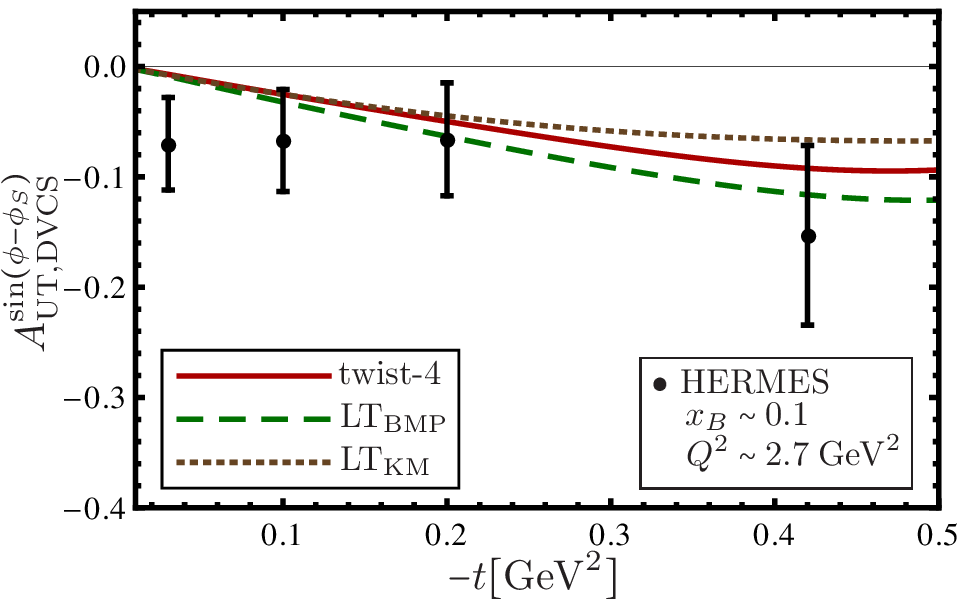}
{}\includegraphics[width=0.5\textwidth]{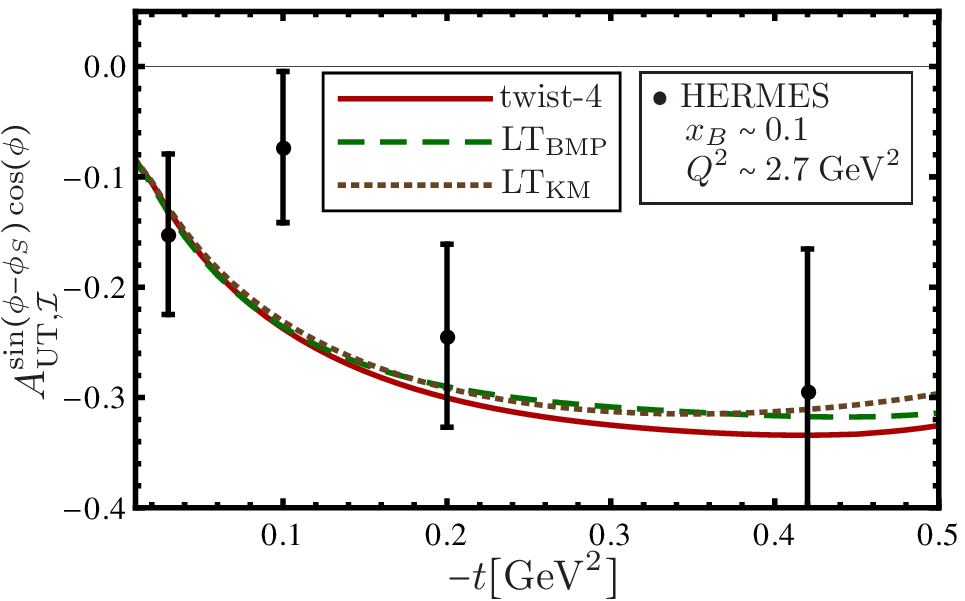}
\vspace{-1mm}
\caption{Transverse target spin asymmetries by HERMES collaboration~\cite{Airapetian:2008aa}. 
%Identification of the curves follows that in Fig.~1.
}
\label{FIG:TRANS}
\end{figure}
The similar comparison for several DVCS observables for transversely polarized targets that are key to access the 
GPD $E$, is done in Ref.~\cite{BjoernThesis}, see 
an example in Fig.~\ref{FIG:TRANS}.
Surprisingly,  we observe that also here the twist-corrections are rather mild. 
However, it is also known  that the model dependence is rather strong~\cite{Kumericki:2013br}.

To summarize, we have carried out a detailed numerical analysis of finite-$t$ and target mass corrections in DVCS,
based on the recent calculation~\cite{Braun:2011zr,Braun:2011dg,Braun:2012bg,Braun:2012hq} 
of the DVCS tensor to twist-four accuracy taking into account the descendants of the leading-twist operators. 
In order to discuss the impact of kinematic higher-twist corrections
one has to formulate the LT approximation that would serve as the reference. This
choice is not unique as the LT calculations are intrinsically ambiguous.
In particular the change in the definition of the
skewedness parameter has a large effect. It turns out that at
least for some observables this difference presents the main
source (numerically) of kinematic corrections, whereas the
remaining higher-twist contributions to the BMP CFFs are
rather mild.
In future phenomenological studies it is highly advisable
to implement besides the kinematical corrections also
perturbative next-to-leading order corrections and, certainly,
GPD evolution must be taken properly into account. This requires
a change to global fitting routines that are based on
appropriate GPD model parametrizations.

\vskip0.2cm

\noindent
{\bf Acknowledgments:}~~{This study was supported by the DFG, grant BR2021/5-2.}

%\acknowledgments{This work was supported by the DFG, grant BR2021/5-2.}

\end{document}